# Geometric phases in Quantum Mechanics

Y.BEN-ARYEH

Physics Department , Technion-Israel Institute of Technology ,
Haifa 32000, Israel
Email: phr65yb@physics.technion.ac.il

Various phenomena related to geometric phases in quantum mechanics are reviewed and explained by analyzing some examples. The concepts of 'parallelisms', 'connections' and 'curvatures' are applied to Aharonov-Bohm (AB) effect, to U(1) phase rotation, to $SU(2)$ phase rotation and to holonomic quantum computation (HQC). The connections in Schrodinger equation are treated by two alternative approaches. Implementation of HQC is demonstrated by the use of 'dark states' including detailed calculations with the connections, for implementing quantum gates. 'Anyons' are related to the symmetries of the wave functions, in a two-dimensional space, and the use of this concept is demonstrated by analyzing an example taken from the field of Quantum Hall effects.

## 1. Introduction

There are various books treating advanced topics of geometric phases [1-3]. The solutions for certain problems in this field remain , however, not always clear. The purpose of the present article is to explain various phemomena related to geometric phases and demonstrate the explanations by analyzing some examples.

In a previous Review [4] the use of geometric phases related to $U(1)$ phase rotation [5] has been described. The geometric phases have been related to topological effects [6], and an enormous amount of works has been reviewed, including both theoretical and experimental results. The aim of the present article is to extend the previous treatment [4] so that it will include now more fields , in which concepts of geometric phases are applied. In the present paper the geometric phases are treated by a unified approach relating such effects to basic topological concepts [6].

Although the topic of Aharonov-Bohm (AB) is well known and understood [7], we analyze the relation between this effect and basic topological concepts [6], so that the similarity between this topic and other problems treated in the present article will be obvious. Although the $U(1)$ geometric phase has been treated extensively in the previous Review [4], we would like to show by following the analysis given in [8] two alternative approaches for the definition of the connections. It is emphasized that by using the approach for single-valuedness of the wave function it is possible to obtain *half-integer* values for the orbital angular momentum [4,9].

In recent years the field of quantum computation has become of much interest [10,11]. It has been suggested to implement quantum gates by the use of Non-Abelian topological quantum computation [12]. Implementation of quantum gates by the use of Holonomic Quantum Computation (HQC) has been suggested where information is encoded in degenerate states. By using the Wilczec-Zee connections [13,14] unitary transformations are obtained which implement quantum computation gates [15-24].

The 'Anyons' concept has been related to the symmetries of the wave functions in a two-dimensional space [25-28]. In order to explain this concept we analyze certain Quantum Hall effects which can be related to the concept of Anyons. There are various books [29-31] treating Quantum Hall effects. In the present Review we relate the concept of Anyons to geometric phases.

The present article is arranged as follows:
In Section 2 the relations between AB effect and topological concepts are described. In this Section the generalization of AB effect to $SU(2)$ phase transformation is also analyzed [32]. As our interest in the present article is in geometric phases related to Schrodinger equation we give here a very simple version of Non-Abelian gauge theory [32]. In Section 3 we discuss two alternative approaches for defining the connections for geometric phases obtained by $U(1)$ phase rotation. In Section 4 HQC is discussed with detailed calculations for one example. 'Anyons' are treated in Section 5 and this concept is related to geometric phases and demonstrated by analyzing an example taken from the field of Quantum Hall effects. We summarize our results in Section 6.

## 2. Parallelism, connections and curvatures in Aharonov-Bohm (AB) effect and its generalization to SU(2) phase rotation

The concept of 'geometric phase' is related to some analogies. The geometric phase analyzed for the Schrodinger equation (the Berry phase[5]) is analogous to the geometric phase obtained for electrons surrounding solenoid with magnetic field (the AB effect [7]; see also [33]). In a similar way there is an analogy between the geometric phase obtained in $SU(2)$ phase rotation [32] and the geometric phase obtained for degenerate states of Schrodinger equation, which is the basis of HQC [15-24]. Although our aim is to explain general topics of geometric phases for systems satisfying Schrodinger equation it will be instructive first to describe the concept of 'geometric phase' in relation to AB effect and its generalization to $SU(2)$ phase rotation.

A particle is described by a complex wave function with a phase. Assuming that the wave function is non-degenerate one is allowed to perform $U(1)$ phase rotation without altering any physics. This fact is referred often as 'local gauge invariance' of the theory. One can, however, compare the phases at two different points of the particle trajectory using the concept of 'parallelism'. Let us explain this concept in AB effect. The phase of the wave function representing a particle of charge $e$ at a spatial point $\vec{x}$ is *parallel* to the phase at a point $\vec{x} + d\vec{x}$ if the local values of the phases at these two points differ by an amount $ea_\mu(\vec{x})dx^\mu$. Here $\vec{a}(\vec{x})$ is the electro-magnetic vector potential at point $\vec{x}$ and $a_\mu(\vec{x})$ is the derivative of this vector potential according to the coordinate $x_\mu$. The wave function at point $\vec{x} + d\vec{x}$ is changed as

$$\Psi(\vec{x}+d\vec{x}) = \Psi(\vec{x}) + ea_\mu(\vec{x})dx^\mu \quad , \quad \mu = 1,2,3 \quad . \quad (2.1)$$

Eq. (2.1) is referred as the *connection* and it is based on the above concept of *parallelism* including the use of vector potential. But the vector potential is defined only up to a *gauge transformation* and as we show now that such arbitrariness is eliminated only for a closed circuit.

Suppose we perform 'a gauge transformation', i.e., rotating the phase of the wave function by an amount $e\alpha(\vec{x})$ so that

$$\Psi(\vec{x}) \to \Psi'(\vec{x}) = \exp(ie\alpha(\vec{x}))\Psi(\vec{x}) \quad . \quad (2.2)$$

The phase at neighboring point will be changed as

$$\Psi(\vec{x}+d\vec{x}) \to \Psi'(\vec{x}+d\vec{x}) = \exp[ie(\alpha(\vec{x})+\partial_\mu\alpha(\vec{x})dx^\mu)]\Psi(\vec{x}+d\vec{x}) \quad . \quad (2.3)$$

In order to get *parallelism between two neighboring* points for $\Psi'(\vec{x})$ the vector potential $a'_\mu(\vec{x})$ in the transformed system is given as

$$a'_\mu(\vec{x}) = a_\mu(\vec{x}) + \partial_\mu\alpha(\vec{x}) \quad . \quad (2.4)$$

One should notice that although $a_\mu(\vec{x})$ is to some extent arbitrary, due to the gauge relation (2.4), this arbitrariness vanish for a closed circle as

$$\oint \partial_\mu\alpha(\vec{x})dx^\mu = 0 \quad . \quad (2.5)$$

For a parallel transport of the wavefunction over a closed circuit the phase will be changed by

$$\phi(C) = e\oint a_\mu(\vec{x})dx^\mu = e\iint_S f_{\mu\nu} d\sigma^{\mu\nu} \quad , \qquad (2.6)$$

where

$$f_{\mu\nu}(\vec{x}) = \partial_\nu a_\mu(\vec{x}) - \partial_\mu a_\nu(\vec{x}) \quad , \qquad (2.7)$$

and (2.6) is obtained by Stokes theorem. $f_{\mu\nu}(\vec{x})$ is known as the EM field tensor and in topology it is is referred as the *curvature*. The phase obtained by (2.6) represents the change of phase obtained by a *parallel transport* along *a closed curve C* in which the particle returns to its initial location. The integration $e\iint_S f_{\mu\nu} d\sigma^{\mu\nu}$ is over the surface bounded by the closed circuit. The phase given by (2.6) will vanish if the vector function $\vec{a}(\vec{x})$ will be curl free, i.e., if $f_{\mu\nu} = 0$ at every point on the *surface* bounded by the closed circuit.

In a generalization of AB effect [32] the wavefunction has two components, i.e., $\Psi(\vec{x}) \to \Psi^i(\vec{x})$ $(i = 1, 2)$. By a change of phase we mean the transformation

$$\vec{\Psi}(\vec{x}) \to \vec{\Psi}'(\vec{x}) = S(\vec{x})\vec{\Psi}(\vec{x}) \quad , \qquad (2.8)$$

where $S$ is a $2 \times 2$ unitary matrix with a unit determinant. The *local gauge invariance* is now the requirement that physics is unchanged under arbitrary $SU(2)$ transformation. The phase at a point $\vec{x}$ is to be regarded to be *parallel* to the phase at a neighboring point $\vec{x} + d\vec{x}$ if the local values of the 'phase' at these two points differ by the amount $gA_\mu(\vec{x})dx^\mu$, for particle with charge $g$, where A is a $2 \times 2$ matrix. Due to the fact that the phase changes are given by matrices which generally do not commute (known in group theory as Non-Abelian) the formulae deduced above for the EM field should be modified.

Before using the transformation (2.8) $\vec{\Psi}(\vec{x})$ was *parallel* to $\vec{\Psi}(\vec{x}+d\vec{x})$ under the condition

$$\vec{\Psi}(\vec{x}+d\vec{x}) = \exp[ig A_\mu(\vec{x})dx^\mu]\vec{\Psi}(\vec{x}) \qquad . \tag{2.9}$$

The explicit evaluation of the generalized potentials $A_\mu$, given by $2\times 2$ matrices, depends on the special physical system. Such *parallelism* will be related later to Wilczec-Zee connections [13,14], for degenerate states of Schrodinger equation. Although we treat here two-dimensional connections the generalization to any $n \geq 2$ connection is straight forward.

After using the transformation (2.8) $\vec{\Psi}(\vec{x})$ is changed to $\vec{\Psi}'(\vec{x})$, while at $\vec{x}+d\vec{x}$ we get

$$\vec{\Psi}'(\vec{x}+d\vec{x}) = S(\vec{x}+d\vec{x})\vec{\Psi}(\vec{x}+d\vec{x}) \qquad . \tag{2.10}$$

In the frame denoted by prime *parallelism* is defined by using potentials $A'_\mu(\vec{x})$ as

$$\vec{\Psi}'(\vec{x}+d\vec{x}) = \exp[ig A'_\mu(\vec{x})dx^\mu]\vec{\Psi}'(\vec{x}) \qquad . \tag{2.11}$$

Substituting (2.9) into the right side of (2.10), and (2.8) into the right side of (2.11) and equating these two equations we get

$$S(\vec{x}+d\vec{x})\exp[ig A_\mu(\vec{x})dx^\mu]\vec{\Psi}(\vec{x}) = \exp[ig A'_\mu(\vec{x})dx^\mu]S(\vec{x})\vec{\Psi}(\vec{x}). \tag{2.12}$$

Since Eq. (2.12) is valid for all $\vec{\Psi}(\vec{x})$ we can omit $\vec{\Psi}(\vec{x})$ and by expanding the equation to first order in $dx^\mu$ we get

$$A'_\mu(\vec{x}) = S(\vec{x})A_\mu(x)S^{-1}(\vec{x}) - \frac{i}{g}[\partial_\mu S(\vec{x})]S^{-1}(\vec{x}) \tag{2.13}$$

When $S$ is infinitesimal, i.e, $S(\vec{x}) \cong 1 + ig\Lambda(\vec{x})$, then to first order in $\Lambda$ we get

$$A'_\mu(\vec{x}) = A_\mu(\vec{x}) + \partial_\mu\Lambda(\vec{x}) + ig[\Lambda(\vec{x}), A_\mu(\vec{x})] \qquad . \qquad (2.14)$$

The *path-ordered phase for a closed loop C* is given for the Non-Abelian case as

$$\phi(C) = P\exp[ig\oint_C A_\mu(\vec{x})dx^\mu] \qquad , \qquad (2.15)$$

known as the Wilson-loop [32]. Since in general cases of the Wilson loops the matrices at different points do not commute we should use path ordering in manipulating (2.15) and computation of the Wilson-loops might be quite complicated. One should, however, take into account that the operation of the Wilson-loop on the basis wavefunctions is given as

$$\vec{\Psi}'_C(\vec{x}) = P\exp[ig\oint_C A_\mu(\vec{x})dx^\mu]\vec{\Psi}_{in}(\vec{x}) \qquad , \qquad (2.16)$$

where $\vec{\Psi}_{in}(\vec{x})$ is the initial wave function and $\vec{\Psi}'_C(\vec{x})$ is the wave function of the system obtained after performing a closed loop $C$. The final result for such computation gives a unitary matrix operating on the initial wave functions with a dimension equal to the dimension of the wave functions' basis. Since quantum computation is based on unitary transformation various authors [12,15-24] developed various methods to implement unitary quantum gates transformations [10,11] by Wilson-loop transformations, for degenerate states of the Schrodinger equation. The dimension of the unitary matrix will then be equal to the number of degenerate states. Since such methods for implementing quantum gates is based on closed loop calculations they are referred as Holonomic Quantum Computation (HQC). Although we will define now the concept of *curvatures* for the Non-Abelian case, one should take into account that in the general case in which the matrices $A_\mu$ at different points do not commute, Stokes theorem *cannot be used for calculating Wilson-loops.*

In a similar way to the Abelian case, $F_{\mu\nu}(\vec{x})$ is defined for the Non-Abelian case to be such that $gF_{\mu\nu}(\vec{x})dx^\mu dx^\nu$ is the difference in 'phase' obtained by parallel transport along the two infinitesimal paths $dx^\mu dx^\nu$ and $dx^\nu dx^\mu$ bringing the initial point to the same final point, where these two infinitesimal paths produce infinitesimal parallelogram. $F_{\mu\nu}(\vec{x})$ is known in topology again as the *curvature*. A straight forward calculation for the Non-Abelian curvature gives [6,32]:

$$F_{\mu\nu}(\vec{x}) = \partial_\nu A_\mu(\vec{x}) - \partial_\mu A_\nu(\vec{x}) + ig[A_\mu(\vec{x}), A_\nu(\vec{x})] \qquad . \qquad (2.17)$$

Eq.(2.17) is reduced to (2.7) under the condition $[A_\mu(\vec{x}), A_\nu(\vec{x})] = 0$. By the transformation made by $S(\vec{x})$ operating on $\vec{\Psi}(\vec{x})$ one finds [6,32] that $F_{\mu\nu}(\vec{x})$ is transformed as

$$F'_{\mu\nu}(\vec{x}) = S(\vec{x})F_{\mu\nu}(\vec{x})S^{-1}(\vec{x}) \qquad . \qquad (2.18)$$

## 3. Geometric phases obtained by U(1) phase rotation for physical systems satisfying Schrodinger equation

The density operator of a pure state satisfying Schrodinger equation is given by

$$\left|\hat{\rho}(\vec{x},t)\right\rangle = \left|\psi(\vec{x},t)\right\rangle\left\langle\psi(\vec{x},t)\right| \qquad , \qquad (3.1)$$

where $\vec{x}$ defines a set of coordinates $x^\mu$. The ket $\left|\Psi(\vec{x},t)\right\rangle$ is determined by $\hat{\rho}$ only up to a phase factor and this phase may depend on the path in which the particle is traversing. We can define $\left|\Psi(\vec{x},t)\right\rangle$ as [8]:

$$\left|\Psi(\vec{x},t)\right\rangle = \left|\chi(\vec{x},t)\right\rangle \exp(i\lambda(\vec{x},t)) \qquad (3.2)$$

where $\exp(i\lambda(\vec{x},t))$ is a phase factor and we impose the phase convention that $\left|\chi(\vec{x},t)\right\rangle$ is a *single-valued function*, returning to its initial value after a closed path it is traversing during a

time T. Such definition can be related in topology [6] to the concept of 'fibre-bundle' where $|\chi(\vec{x},t)\rangle$ is the basis of the fiber bundle and $\exp(i\lambda(\vec{x},t))$ is the fiber.

For infinitesimal change of the ket $|\Psi(\vec{x},t)\rangle$ we get

$$|d\Psi\rangle = |d\chi\rangle\exp(i\lambda) + |\chi\rangle id\lambda \exp(i\lambda) \quad . \tag{3.3}$$

Multiplying (3.3) on the left by $\langle\Psi|$ we get

$$\langle\Psi|d\Psi\rangle = \langle\chi|d\chi\rangle + id\lambda \quad . \tag{3.4}$$

Due to normalization condition we have

$$\langle\Psi|d\Psi\rangle + \langle d\Psi|\Psi\rangle = \langle\chi|d\chi\rangle + \langle d\chi|\chi\rangle = 0 \quad . \tag{3.5}$$

We get from (3.4)

$$d\lambda = \frac{1}{i}\left(\langle\Psi|d\Psi\rangle - \langle\chi|d\chi\rangle\right) \quad . \tag{3.6}$$

Integrating (3.6) around the closed circuit $C$ and using relations (3.5) the phase change is obtained as [8]:

$$\Delta\lambda(C) = f(C) - F(C) \quad ; \tag{3.7}$$

$$f(C) = \frac{1}{i}\oint_C \langle\Psi|d\Psi\rangle = \frac{1}{2i}\oint_C \left(\langle\Psi|d\Psi\rangle - \langle d\Psi|\Psi\rangle\right) \quad ; \tag{3.8}$$

$$F(C) = \frac{1}{i}\oint_C \langle\chi|d\chi\rangle = \frac{1}{2i}\oint_C \left(\langle\chi|d\chi\rangle - \langle d\chi|\chi\rangle\right) \quad . \tag{3.9}$$

The case of $U(1)$ phase rotation is referred as the Abelian case, since the integrands in (3.8-9) at different points along the curve commute with one another.

Time development under the Schrodinger equation leads to the relation

$$|d\Psi(t)\rangle = -iH(t)|\Psi(t)\rangle \quad , \tag{3.10}$$

where $H(t)$ is the Hamiltonian operator. Multiplying (3.10) on the left by $\langle \Psi(t)|$, integrating over the closed circuit during a time T and using (3.8) we get

$$f(C) = -\int_0^T \langle \Psi(t)|H(t)|\Psi(t)\rangle dt \qquad . \tag{3.11}$$

$f(C)$ is equal to minus expectation value of the energy $E$ over time. The contribution of $f(C)$ to the final change in phase $\Delta\lambda(C)$ is referred as the 'dynamical phase' given by $\int_0^T E dt$.

The additional term of (3.9) which depends on the path in the projective space is referred as the geometric phase. We can eliminate the dynamical phase by defining [34] :

$$|\tilde{\Psi}(t)\rangle = |\Psi(t)\rangle \exp\left[i\int_0^t \langle \Psi(t')|H(t')|\Psi(t')\rangle dt'\right] \qquad . \tag{3.12}$$

Then one finds that $f(C)$ for $|\tilde{\Psi}(t)\rangle$ vanishes.

There are two approaches for calculating geometric phases which usually give equivalent results. One can introduce the concept of 'parallelism' in Schrodinger equation by which

$$\langle \tilde{\Psi}|d\tilde{\Psi}\rangle = 0 \qquad . \tag{3.13}$$

This condition is obtained in Schrodinger equation by requiring that $|\tilde{\Psi}(t+\Delta t)\rangle$ is in phase with $|\tilde{\Psi}(t)\rangle$, which means that $\langle \tilde{\Psi}(t)|\tilde{\Psi}(t+\Delta t)\rangle$ is real and positive. Since $\langle \tilde{\Psi}(t)|\tilde{\Psi}(t)\rangle = 1$ and $\langle \tilde{\Psi}(t)|d\tilde{\Psi}(t)\rangle$ is necessarily imaginary, Eq. (3.13) follows from the use of 'parallelism' in Schrodinger equation. The connection (3.13) is similar to the connection used by Pancharatnam quite long ago [35], who defined two light beams to be in phase if the intensity of their sum is maximal. The basic idea is that phases do not follow the transitive law so that if system $A$ is in phase with system $B$ and system $B$ is in phase with system $C$ in general cases it does not follow that system $A$ is in phase with system $C$. This idea is applied for the connection (3.13) as *locally* the phase change is vanishing but *globally* on a closed circuit it leads to the geometric phase contribution. Pancharatnam connection which is analogous to (3.13) has been used for calculating geometric phases also for open circuits, and we refer to the literature for such applications [36].

In another approach assuming that $|\Psi(\vec{x},t)\rangle$ depend on set of parameters $\vec{R}$ e.g. $(R_1, R_2, R_3)$ [5] which are functions of time so that $|\Psi(\vec{x},t)\rangle = |\Psi(\vec{R}(t))\rangle$ then the geometric phase can be calculated by

$$F(C) = \frac{1}{i} \oint_C \langle \Psi | \vec{\nabla}_R \Psi \rangle \cdot d\vec{R} \qquad . \qquad (3.14)$$

Eq. (3.14) can be generalized to any number $n$ of parameteric coordinates $R_1, R_2 \cdots R_n$. The use of (3.14) is analogous to the calculation of the AB phase in Eq. (2.6). The coordinates $\vec{x}$ of (2.6) are replaced in (3.14) by coordinates $\vec{R}$. The derivative of the vector potential in Eq. (2.6) is replaced in (3.14) by the vector potential derivative $\langle \Psi | \vec{\nabla}_R \Psi \rangle$. For the $U(1)$ Abelian case of Schrodinger equation, $F(C)$ can be calculated by Stokes theorem in analogous way to the calculation given by (2.6) for the AB effect.

One should notice that in the second approach 'parallelism' can be defined in an analogy to AB effect by

$$\Delta \lambda = \langle \Psi | \vec{\nabla}_R \Psi \rangle \cdot d\vec{R} \qquad , \qquad (3.15)$$

but this definition is arbitrary up to a gauge transformation analogous to (2.4) in the AB effect. Only for a closed circuit we get a well defined geometric phase independent of gauge transformation.

The following example taken from Ref. [8] illuminates the difference between the two approaches. Let us assume an Hamiltonian in a two-dimensional space of the form

$$H = \begin{pmatrix} x & y \\ y & -x \end{pmatrix} = r \begin{pmatrix} \cos\phi & \sin\phi \\ \sin\phi & -\cos\phi \end{pmatrix} . \qquad (3.16)$$

The eigenvalues of this Hamiltonian are

$$|\chi_+\rangle = \begin{pmatrix} \cos\frac{\phi}{2} \\ \sin\frac{\phi}{2} \end{pmatrix}, \quad |\chi_-\rangle = \begin{pmatrix} -\sin\frac{\phi}{2} \\ \cos\frac{\phi}{2} \end{pmatrix} . \qquad (3.17)$$

Under such conditions

$$|\delta\chi_+\rangle = \frac{1}{2}|\chi_-\rangle \delta\phi \; ; \; |\delta\chi_-\rangle = -\frac{1}{2}|\chi_+\rangle \delta\phi , \qquad (3.18)$$

so that

$$\langle \vec{\chi} | \delta\vec{\chi} \rangle = \langle \vec{\chi} | \nabla_\phi \vec{\chi} \rangle \cdot \delta\phi = 0 , \qquad (3.19)$$

and Eq. (3.13) is satisfied. We can manipulate the phase factor in such a way that locally the connection (3.19) is satisfied. However as the phases *are not added transitively* there is a global phase change under a closed circuit. In the above approach the eigenkets are not single-valued, however, undergoing a change of sign when $\phi$ makes a full circuit from 0 to $2\pi$. It is quite easy to find in this case a phase factor that will make the kets single-valued. For $|\chi_-\rangle$ for example make the transformation

$$|X_-\rangle = |\chi_-\rangle \exp(i\phi/2) . \qquad (3.20)$$

This transformation restores the single-valuedness, but now there is a non-zero vector potential

$$A_\phi = \frac{1}{i}\langle X_- | \frac{d}{d\phi} X_- \rangle = \frac{1}{2} , \qquad (3.21)$$

so that $A_\phi d\phi$ does not vanish. This example shows that for certain calculations of geometric phases one has to choose between the approach of making the vector potential orthogonal to the movement $d\vec{R}$ leading to the connection

$$\vec{A}(\vec{R}) \cdot d\vec{R} = 0 \qquad , \qquad (3.22)$$

and the approach of single valuedness of the $U(1)$ system, leading to calculations with vector potentials for which $\vec{A}(\vec{R}) \cdot d\vec{R} \neq 0$, like that made in (3.21). Since the vector potential is defined up to relative constant we can always satisfy the connection (3.22). In the vector-potential approach the contribution of the gauge transformation (2.4) vanishes only for a closed circuit so that also in this approach the geometric phase has only a global approach. The vector-potential approach has also the advantage that Stokes theorem can be used for a closed circuit for the Abelian case of $U(1)$ phase rotation. The requirement that the total wave function will be single–valued can lead to interesting effects ,e.g. obtaining half–integer orbital angular momentum in Jahn-Teller system (see e.g. [4,9]). The implication of the single-valuedness of the wave function has been discussed also for a particle with two dimensional Hamiltonian in polar coordinates including harmonic potential function , in relation to 'bosonic' ,'fermionic' and 'anyonic' particles [25]. Half integer orbital angular momentum occurs also for non-Abelian fields [37].

Most experimental and theoretical work on $U(1)$ phase rotation are related to geometric phases in two-level systems and Lorentz-group topological phases. However, such phases have been studied also for three-level atomic systems [38] including their relations with group theory [39].

# 4. Holonomic quantum computation (HQC)

In order to understand the idea of using quantum *holonomic quantum computation* ($HQC$), let us review first some fundamental properties of *quantum computation* ($QC$) [10,11]. The basic unit in $QC$ is the *quantum bit* (qubit) which is a two-level quantum system described by a two-dimensional complex Hilbert space. Matrix representations of the two levels defined as qubits $|0\rangle$ and $|1\rangle$ are given by the following column vectors

$$|0\rangle \equiv \begin{pmatrix} 1 \\ 0 \end{pmatrix} \ ; \ |1\rangle \equiv \begin{pmatrix} 0 \\ 1 \end{pmatrix} . \tag{4.1}$$

$QC$ is based on the use of certain basic unitary operations defined as *quantum gates*. The unitary transformations operating on these single qubit column vectors can be given by multiplying them by unitary matrices of dimension $2 \times 2$:

$$I = \begin{pmatrix} 1 & 0 \\ 0 & 1 \end{pmatrix} \ , \ \sigma_1 = \begin{pmatrix} 0 & 1 \\ 1 & 0 \end{pmatrix} \ , \ \sigma_2 = \begin{pmatrix} 0 & -i \\ i & 0 \end{pmatrix} \ , \ \sigma_3 = \begin{pmatrix} 1 & 0 \\ 0 & -1 \end{pmatrix} \ , \tag{4.2}$$

where $I$ is the two-dimensional unit matrix, and $\sigma_1, \sigma_2, \sigma_3$ are the Pauli spin matrices. Single qubit gates are based on unitary transformations in which Pauli spin matrices are used.

The two-qubit states can be given by four dimensional column vectors:

$$|00\rangle \equiv \begin{pmatrix} 1 \\ 0 \end{pmatrix} \otimes \begin{pmatrix} 1 \\ 0 \end{pmatrix} \equiv \begin{pmatrix} 1 \\ 0 \\ 0 \\ 0 \end{pmatrix} \ ; \ |01\rangle \equiv \begin{pmatrix} 1 \\ 0 \end{pmatrix} \otimes \begin{pmatrix} 0 \\ 1 \end{pmatrix} \equiv \begin{pmatrix} 0 \\ 1 \\ 0 \\ 0 \end{pmatrix}$$

$$|10\rangle \equiv \begin{pmatrix} 0 \\ 1 \end{pmatrix} \otimes \begin{pmatrix} 1 \\ 0 \end{pmatrix} \equiv \begin{pmatrix} 0 \\ 0 \\ 1 \\ 0 \end{pmatrix} \ ; \ |11\rangle \equiv \begin{pmatrix} 0 \\ 1 \end{pmatrix} \otimes \begin{pmatrix} 0 \\ 1 \end{pmatrix} \equiv \begin{pmatrix} 0 \\ 0 \\ 0 \\ 1 \end{pmatrix} \tag{4.3}$$

where in the ket states on the left handside of these equations the first and second number denote the state of the first and second qubit, respectively. The sign $\otimes$ represents tensor product where the two-qubit states can be described by tensor products of the first and second qubit column vectors.

Any two-qubit gate operating on the two qubit states can be given by a unitary matrix $U_2$ of dimension $4 \times 4$ multiplying the 4 dimensional vectors describing the two-qubit states by (4.3). The $CNOT$ gate is operating on the four dimensional vectors of (4.3) by the unitary matrix

$$CNOT = \begin{pmatrix} 1 & 0 & 0 & 0 \\ 0 & 1 & 0 & 0 \\ 0 & 0 & 0 & 1 \\ 0 & 0 & 1 & 0 \end{pmatrix}. \qquad (4.4)$$

The SWAP gate is operating on the four dimensional vectors of (4.3) by the unitary matrix

$$SWAP = \begin{pmatrix} 1 & 0 & 0 & 0 \\ 0 & 0 & 1 & 0 \\ 0 & 1 & 0 & 0 \\ 0 & 0 & 0 & 1 \end{pmatrix}. \qquad (4.5)$$

The phase gate is operating on the four dimensional vectors of (4.3) by

$$PHASE = \begin{pmatrix} 1 & 0 & 0 & 0 \\ 0 & 1 & 0 & 0 \\ 0 & 0 & 1 & 0 \\ 0 & 0 & 0 & \exp(i\phi) \end{pmatrix}. \qquad (4.6)$$

The experimental realization of the quantum gates, their properties and applications in $QC$ has described by many methods in an enormous amount of works [10,11 (see also [40] and References included). In the present Section we would like to discuss, however, the use of $HQC$.

According to basic ideas of *HQC* information is encoded in a degenerate eigenspace of an Hamiltonian $H(\vec{\lambda})$ which is a function of parameters $\vec{\lambda} = (\lambda_1, \lambda_2 \cdots \lambda_d)$ and these parameters are varying slowly so that the evolution is adiabatic ,i.e., no transitions are induced among different energy levels. Let us assume that we have an orthonormal basis of states $\{|\psi^\alpha(\vec{\lambda})\rangle\}$, $\alpha = 1,2 \cdots n$ in a degenerate space $\Gamma$. An initial state $|\psi_0\rangle \in \Gamma$ is developed in an adiabatic way by a slow change of the control parameters $\vec{\lambda}$ where $\vec{\lambda}$ has $d$ components denoted as $\vec{\lambda}_\mu$, $\mu = 1,2 \cdots d$. Assuming that the energy of the degenerate state is equal to zero, the dynamical phase vanishes and the initial state is developed by the unitary transformation

$$|\psi_0\rangle \to U(\vec{\lambda})|\psi_0\rangle \quad ; \quad U(\vec{\lambda}) = \text{P} \exp \oint A = \text{P} \exp\left(-i \oint \sum_{\mu=1}^{d} A_\mu d\lambda_\mu \right) \quad , (4.7)$$

where $A_\mu$ is a matrix of $n \times n$ dimension, known as the Wilczec-Zee connection [13,14] , with matrix elements given by

$$A_\mu^{(\alpha,\beta)} = \left\{ \langle \psi^\alpha(\vec{\lambda}) | \frac{\partial}{\partial \lambda_\mu} | \psi^\beta(\vec{\lambda}) \rangle \right\} \quad , \quad \alpha, \beta = 1,2 \cdots n \qquad (4.8)$$

and where P denotes path ordering . The calculation of $U(\vec{\lambda})$ for a closed circuit in the $\vec{\lambda}$ parameters is known as 'holonomy'. In *HQC* we would like to use such holonomies for implementing quantum gates.

While the realization of single-qubit gates is relatively simple the use of two-qubit gates which are based on controlled operations is quite complicated and we would like to demonstrate here the detailed calculations in an example for two-qubit gates following the analysis made in [18].

Let us consider the $(5+1)$ system [18] which under zeroth order Hamiltonian $H_0$ has five eigenvalues equal to zero and correspondingly five orthonormal eigenvectors

$$|g_1\rangle = \begin{pmatrix} 0 \\ 1 \\ 0 \\ 0 \\ 0 \\ 0 \end{pmatrix} \; ; \; |g_2\rangle = \begin{pmatrix} 0 \\ 0 \\ 1 \\ 0 \\ 0 \\ 0 \end{pmatrix} \; ; \; |g_3\rangle = \begin{pmatrix} 0 \\ 0 \\ 0 \\ 1 \\ 0 \\ 0 \end{pmatrix} \; ; \; |g_4\rangle = \begin{pmatrix} 0 \\ 0 \\ 0 \\ 0 \\ 1 \\ 0 \end{pmatrix} \; ; \; |g_5\rangle = \begin{pmatrix} 0 \\ 0 \\ 0 \\ 0 \\ 0 \\ 1 \end{pmatrix} \; ; \; |e\rangle = \begin{pmatrix} 1 \\ 0 \\ 0 \\ 0 \\ 0 \\ 0 \end{pmatrix} ,(4.9)$$

and we include in the present system also an excited state with energy $\varepsilon$ and eigenvector $|e\rangle$. The six levels are defined so that they satisfy the orthonormal relations:

$$\langle g_i | g_k \rangle = \delta_{i,k} , (i,k = 1,2\cdots 5) \;\;,\;\; \langle e | e \rangle = 1 \;\;,\;\; \langle g_i | e \rangle = 0, \; (i = 1,2\cdots 5). \quad (4.10)$$

Let assume that we add an Hamiltonian of interaction

$$H_1 = \sum_{k=1}^{5} \Omega_k |g_k\rangle\langle e| + H.C. \qquad (4.11)$$

by which five ground states $|g_k\rangle$, each with zero energy, are coupled to the excited level $|e\rangle$ which has energy $\varepsilon$. The complex coupling constants $\Omega_k$ replace the parameters $\vec{\lambda}$. There are no terms coupling directly the ground states, but they are coupled to the upper state.

The Hamiltonian for the $(5+1)$ system is then given as

$$H = \begin{pmatrix} \varepsilon & \Omega_1^* & \Omega_2^* & \Omega_3^* & \Omega_4^* & \Omega_5^* \\ \Omega_1 & 0 & 0 & 0 & 0 & 0 \\ \Omega_2 & 0 & 0 & 0 & 0 & 0 \\ \Omega_3 & 0 & 0 & 0 & 0 & 0 \\ \Omega_4 & 0 & 0 & 0 & 0 & 0 \\ \Omega_5 & 0 & 0 & 0 & 0 & 0 \end{pmatrix} . \qquad (4.12)$$

Here the coupling constants can be written by some kind of spherical coordinates [18]:

$$\begin{aligned}
\Omega_1 &= |\Omega| \sin\theta_1 \;;\quad \Omega_2 = |\Omega| \cos\theta_1 \sin\theta_2 \exp(-i\phi_2) \;; \\
\Omega_3 &= |\Omega| \cos\theta_1 \cos\theta_2 \sin(\theta_3) \exp(-i\phi_3) \;; \\
\Omega_4 &= |\Omega| \cos\theta_1 \cos\theta_2 \cos(\theta_3) \sin(\theta_4) \exp(-i\phi_4) \;; \\
\Omega_5 &= |\Omega| \cos\theta_1 \cos\theta_2 \cos(\theta_3) \cos(\theta_4) \exp(-i\phi_5) \;; \\
|\Omega|^2 &= |\Omega_1|^2 + |\Omega_2|^2 + |\Omega_3|^2 + |\Omega_4|^2 + |\Omega_5|^2 \;.
\end{aligned} \qquad (4.13)$$

so that now $\theta_1, \theta_2, \theta_3, \theta_4, \phi_2, \phi_3, \phi_4$ and $\phi_5$ replace the parameters $\vec{\lambda}$. In [18] the four states $|\Psi^1\rangle$, $|\Psi^2\rangle$, $|\Psi^3\rangle$, $|\Psi^4\rangle$ which are eigenstates of $H$ with zero eigenvalues and which compose the degenerate eigenspace $\Gamma$ have been calculated. Such states are referred often as *dark states* [18,41,42], i.e., zero energy states having no contribution from the excited state. In [18] it is assumd that the system is developed adiabatically within these degenerate eigenstates where the Wilczec–Zee connections can be calculated by Eq.(4.8) replacing $\vec{\lambda}_\mu$ by $\Omega_k$, $k = 1,2,3,4,5$, or explicitly by $\theta_1, \theta_2, \theta_3, \theta_4, \phi_2, \phi_3, \phi_4$ and $\phi_5$ and where $\alpha$ and $\beta$ go over the wave functions $|\Psi^\alpha\rangle$ and $|\Psi^\beta\rangle$ $(\alpha, \beta = 1,2,3,4)$. Although the Hamiltonian (4.12) admits two excited states with energies $E_+ = \frac{\varepsilon}{2} + \sqrt{\left(\frac{\varepsilon}{2}\right)^2 + |\Omega|^2}$ and $E_- = \frac{\varepsilon}{2} - \sqrt{\left(\frac{\varepsilon}{2}\right)^2 + |\Omega|^2}$,

we assume that our system develops within the lower 'dark states' and the excited states are not relevant to the present $HQC$.

It has been shown in [18] how to use the above system to implement the *PHASE* gate. They have also shown how to use the (3+1) 'dark states' system known as the 'Λ system' for implementing any single qubit gate [18]. We refer to this work for these implementations. Here we show how to use this system for implementing the *CNOT* gate which has not been discussed in [18].

A simple analysis can be obtained from the above complicated system by fixing

$$\theta_1 = \theta_2 = \phi_2 = \phi_3 = \phi_4 = \phi_5 = 0 \rightarrow \Omega_1 = \Omega_2 = 0 \; ; \; \Omega_3 = |\Omega_3| \; ; \; \Omega_4 = |\Omega_4| ; \; \Omega_5 = |\Omega_5| \quad (4.14)$$

Under these conditions the four eigenfunctions of the Hamiltonian (4.12) with zero eigenvalues are given as

$$|\Psi^1\rangle = \begin{pmatrix} 0 \\ 1 \\ 0 \\ 0 \\ 0 \\ 0 \end{pmatrix} \; ; \; |\Psi^2\rangle = \begin{pmatrix} 0 \\ 0 \\ 1 \\ 0 \\ 0 \\ 0 \end{pmatrix} \; ; \; |\Psi^3\rangle = \begin{pmatrix} 0 \\ 0 \\ 0 \\ \cos\theta_3 \\ -\sin\theta_3 \sin\theta_4 \\ -\sin\theta_3 \cos\theta_4 \end{pmatrix} \; ; \; |\Psi^4\rangle = \begin{pmatrix} 0 \\ 0 \\ 0 \\ 0 \\ -\cos\theta_4 \\ \sin\theta_4 \end{pmatrix} . \quad (4.15)$$

It is easy to verify that the states given by (4.15) are eigenstates of the Hamiltonian (4.12) under the conditions (4.14) with zero eigenvalues. It is also easy to verify that the states (4.15) are orthonormal. The derivatives of $|\Psi^1\rangle$ and $|\Psi^2\rangle$ over any parameter vanish and the derivatives of the functions $|\Psi^3\rangle$ and $|\Psi^4\rangle$ remain only over the parameters $\theta_3$ and $\theta_4$. We get:

$$\frac{\partial \Psi^4}{\partial \theta_4} = \begin{pmatrix} 0 \\ 0 \\ 0 \\ 0 \\ \sin\theta_4 \\ \cos\theta_4 \end{pmatrix} \; ; \; \frac{\partial \Psi^3}{\partial \theta_4} = \begin{pmatrix} 0 \\ 0 \\ 0 \\ 0 \\ -\sin\theta_3 \cos\theta_4 \\ \sin\theta_3 \sin\theta_4 \end{pmatrix} \; ; \; \frac{\partial \Psi^3}{\partial \theta_3} = \begin{pmatrix} 0 \\ 0 \\ 0 \\ -\sin\theta_3 \\ -\cos\theta_3 \sin\theta_4 \\ -\cos\theta_3 \cos\theta_4 \end{pmatrix} . \quad (4.16)$$

By straight forward calculations of the connections (4.8) we obtain the vanishing connections:

$$\left\langle \Psi^3 \mid \frac{\partial \Psi^3}{\partial \theta_3} \right\rangle = -\sin\theta_3 \cos\theta_3 + \sin\theta_3 \cos\theta_3 (\sin^2\theta_4 + \cos^2\theta_4) = 0 \;;$$

$$\left\langle \Psi^3 \mid \frac{\partial \Psi^3}{\partial \theta_4} \right\rangle = -\sin^2\theta_3 \sin\theta_4 \cos\theta_4 + \sin^2\theta_3 \sin\theta_4 \cos\theta_4 = 0 \;;$$

$$\left\langle \Psi^4 \mid \frac{\partial \Psi^4}{\partial \theta_4} \right\rangle = -\sin\theta_4 \cos\theta_4 + \sin\theta_4 \cos\theta_4 = 0 \;;$$

$$\left\langle \Psi^4 \mid \frac{\partial \Psi^3}{\partial \theta_3} \right\rangle = \cos\theta_4 \cos\theta_3 \sin\theta_4 - \cos\theta_4 \cos\theta_3 \sin\theta_4 = 0 \;.$$

(4.17)

We remain with the non-diagonal connections

$$\left\langle \Psi^3 \mid \frac{\partial \Psi^4}{\partial \theta_4} \right\rangle = -\sin\theta_3 \;;$$

$$\left\langle \Psi^4 \mid \frac{\partial \Psi^3}{\partial \theta_4} \right\rangle = \sin\theta_3 \;.$$

(4.18)

Using relations (4.18) we get the unitary transformation

$$U = \exp\left\{ i \begin{pmatrix} 0 & 0 & 0 & 0 \\ 0 & 0 & 0 & 0 \\ 0 & 0 & 0 & \oint \sin\theta_3 d\theta_4 \\ 0 & 0 & \oint \sin\theta_3 d\theta_4 & 0 \end{pmatrix} \right\}.$$

(4.19)

The result in (4.19) depends on the closed integral. This integral is made along a closed line in the $(\theta_4, \theta_3)$ manifold plane and it can be calculated by Stokes theorem where

$$\frac{\partial \sin\theta_3}{\partial \theta_3} = \cos(\theta_3) \;; \quad \frac{\partial \sin\theta_3}{\partial \theta_4} = 0 \;.$$

(4.20)

Then the integral of (4.19) is transformed to surface integral over the plane closed by the manifold loop obtaining

$$U = \exp\left\{ i \begin{pmatrix} 0 & 0 & 0 & 0 \\ 0 & 0 & 0 & 0 \\ 0 & 0 & 0 & \iint_S \cos\theta_3 d\theta_4 d\theta_3 \\ 0 & 0 & \iint_S \cos\theta_3 d\theta_4 d\theta_3 & 0 \end{pmatrix} \right\} . \quad (4.21)$$

The final result for the unitary operator $U$ depends on the calculation of the surface integral. In the general case we need to use series expansion of (4.21), which will produce mixing of the $\Psi^3$ and $\Psi^4$ states. In order to simulate the $CNOT$ gate we choose the surface integral to be equal to $\pi/2$. Then the series expansion of (4.21) leads to the result

$$U = \begin{pmatrix} 1 & 0 & 0 & 0 \\ 0 & 1 & 0 & 0 \\ 0 & 0 & 0 & i \\ 0 & 0 & i & 0 \end{pmatrix} . \quad (4.22)$$

The unitary operator of (4.22) is equal to the $CNOT$ unitary operator (4.4) up to the change in the phase $i$ for the non-diagonal elements. Quantum gates up to relative phases have been referred as 'mapping' operations and can be used in quantum computation (see [10] pages 320,183). In order to get the ordinary $CNOT$ gate one can eliminate the phase $i$ by using the $PHASE$ gate as analyzed in [18].

The above example has been chosen for demonstrating $HQC$ due to its relative simplicity. Usually calculations by $HQC$ are much more complicated. The calculations in the above example have been reduced to the Abelian case. More general cases include Non-Abelian Wilson integrals which are much more complicated. In some cases it is possible to construct a closed loop from some segments of finite length where for each segment an Abelian integration can be made ,i.e., assuming that the integrands are commuting in each segment. In some papers [19-21] the problem of finding a loop , as short as possible in the control parameters , which will implement the quantum gate has been studied. In these papers [19-21] the relations between *HQC* and topology have been also analyzed. Various experimental schemes have been suggested for implementing *HQC* [15-24]. It has been claimed that since the holonomy depends only on the geometry of the control parameters' loop it is insensitive to the details of the dynamics of the system and therefore *HQC* will be robust against noise from the environment.

In the above example of using dark states for implementing two-qubit gates we demonstrated the detailed calculations for the *CNOT* gate. In principle the dark states can be used for implementing other two-qubit gates We leave it as an excercise how to use the dark states for implementing the *SWAP* gate of (4.5).

## 5. Anyons and geometric Phases

Let us review some fundamental properties for the symmetry of particles [28]. If two identical non-interacting particles are exchanged, their wavefunctions will acquire either a plus (bosons) or a minus (fermions) sign. In three-dimensional space this is the only observed statistical behavior. A particle circulating around an identical one spans a path, that in three-dimensional space can be continuously deformed into a point. In topology [6] one uses the concept by which the circuit and the point are *homotopic*. Two successive exchanges of two particles leads to the original configuration by which the physics describing the particle is not changed. In three-dimensional space it is equivalent to a full circulation of one particle around the other. Thus a single exchange can result in a phase factor $\exp(i\theta)$ whose square should be equal to unity. There are therefore two options, giving $\theta = 0$ for bosons and $\theta = \pi$ for fermions.

In two-dimensional space a closed path $\gamma_1$ in which the first particle circulates around the second particle is distinct from a closed path $\gamma_2$ in which the second particle is external to the circuit traversed by the first particle. While in the latter case the circuit can be deformed continuously into a point, in the first case it cannot be done in the plane, since intersection of the circuit traversed by the first particle with the second particle is not allowed. In topology these two cases are considered as *non-homotopic*.

Electrons, protons, atoms and photons are usually either fermions or bosons even when they are confined to move in a two-dimensional plane. In order to get in two-dimensional space particles which are neither bosons or fermions there should be certain constraints by which a single exchange between two particles leads to a phase factor $\exp(i\theta)$ where $\theta$ is different from 0 and $\pi$ [28]. In order to visualize such phase change [1,43] one can think of composite particles consisting of a small ring of a charge $q$ and magnetic flux $\phi$ pentrating it, perpendicular to the two-dimensional plane. If such particle circulates another such particle then it acquires a phase factor $\exp(i\theta)$ due to the AB effect, and such phase can take any value. This is only a *pictorial representation* and the full analysis should be related to gauge theory. Identical particles acquiring by their exchanges phases which are different from 0 and $\pi$ are referred as Anyons [25-28]. These are particles satisfying a statistics which is different from that of fermions or bosons.

There are various systems in which certain phenomena have been related to the concept of 'Anyons'. Such systems include, among others, theoretical descriptions of Anyons for small atomic Bose-Einstein condensates [44], for bosonic fields interacting with fermions in a Jayness–Cumming model [45] and for superconducting field adjacent to two-dimensional electron gas [46]. However, such effects have been observed mainly in the Quantum Hall effects. The field of Quantum Hall effects has been developed in an enormous amount of works and are reviewed in various books [29-31]. In the following discussion the use of the 'Anyons' geometric phase in quantum Hall effect is demonstrated, only for a certain example, following the analysis made in [47] and [30]. There are general theories of Anyons by which the topological trajectories of many particle system are related to the Braid group [27,28,48]. These theories are very complicated and it has been suggested to apply them for the field of quantum computation. In the meantime, there are not experimental observations related to the Braid group and since these theories are beyond the scope of the present article we will refer only to the literature treating these subjects [27,28,48].

Let us review ,shortly, remarkable properties of the Laughlin's theory of the quantum Hall Effects related to the Landau levels [29-31, 49-51]. Consider electrons confined in the $x-y$ plane and subjected to a constant magnetic field $B\hat{z}$ perpendicular to the plane . Using the symmetric gauge vector potential $\vec{A} = \frac{1}{2}B(x\hat{y} - y\hat{x})$, it is convenient to regard the $x-y$ plain as a complex plane. The degeneracy of the present system has been calculated as (see e.g. Section 1.2 in [30] ) as

$$N(s) = \frac{S}{2\pi l_0^2} \quad , \tag{5.1}$$

where $S$ is the area of the system and

$$l_0 = \left(\frac{\hbar c}{eB}\right)^{1/2} . \tag{5.2}$$

Another important quantity is the dimensionless density of the electrons expressed as the *filling factor* of the landau level

$$\nu = \frac{N}{N(s)} = n_0 S \frac{(2\pi l_0^2)}{S} = 2\pi l_0^2 n_0 \qquad . \qquad (5.3)$$

where $n_0$ is the surface electron density. The typical interaction energy per particle is $e^2/d$ where $d$ is the average distance between the electrons. At high magnetic fields it is small compared to the space $\hbar \omega_C$ between the Landau levels where $\omega_C = \frac{eB}{m^* c}$ ($m^*$ is the effective mass of the electron). Therefore one can ignore the coupling between two Landau levels and all the electrons can stay in the lowest low level [2].

The many-electron wave functions proposed by Laughlin for the $m = 1/\nu$ state is

$$\psi_m = \prod_{\substack{i,k=1 \\ j<k}}^{N_e} (z_i - z_k)^m \prod_{j=1}^{N_e} \exp\left(-\sum_i |z_i|^2 / 4l_0^2\right) \qquad , \qquad (5.4)$$

where $m(N_e - 1) = N(s) - 1$ [30], and the coordinates are given in the complex plane, i.e, $z = x - iy$ corresponding to the coordinates $(x, y)$ in two dimensions. For large $N_e$ one obtains [30]

$$m \cong \frac{S}{2\pi l_0^2 N_e} = \frac{1}{\nu} \quad , \quad \frac{N_e}{S} \cong n_0 \qquad . \qquad (5.5)$$

For the derivation of the above equation and its various properties see [29-31, 49-51]. Here we would like to describe only an example of geometric phase calculation related to the concept of 'Anyons'.

Laughlin wave function for the quasihole located at position $z_0$ has been described as

$$\psi_m^{(-)} = N_- \prod_i (z_i - z_0) \psi_m \tag{5.6}$$

where $\psi_m$ has been given in (4.4), and $N_-$ is a normalization constant. Let the position $z_0$ of the quasihole move adiabatically around a closed circle $C$ in the $(x, y)$ plane, the rate of the phase change can be expressed as

$$\frac{d\gamma(t)}{dt} = i \left\langle \psi_m^{(-)} \middle| \frac{d\psi_m^{(-)}}{dt} \right\rangle = i \left\langle \psi_m^{(-)} \middle| \frac{d}{dt} \ln[z_i - z_0(t)] \middle| \psi_m^{(-)} \right\rangle . \tag{5.7}$$

Using the one particle density

$$\rho^{(-)}(z) = \left\langle \psi_m^{(-)} \middle| \sum_i \delta(z_i - z) \middle| \psi_m^{(-)} \right\rangle , \tag{5.8}$$

of the quasihole state, Eq.(5.7) can be written as

$$\frac{d\gamma(t)}{dt} = i \iint dxdy \, \rho^{(-)}(z) \frac{d}{dt} \ln[z - z_0(t)] , \tag{5.9}$$

where $z = x + iy$. Assuming $\rho^{(-)}(z) \cong \rho_0$, where $\rho_0$ is the averaged density, and that the contribution of the fluctuations to the integral of (5.9) are negligible relative to that of $\rho_0$, we get for integration over a disk of radius $R$ [30,47]:

$$\gamma(T) = i \iint_{|r|<R} dxdy \, \rho_0 \, 2\pi i = -2\pi \langle n \rangle_R = -2\pi \nu \frac{\Phi}{\Phi_0} . \tag{5.10}$$

Here we have used the relation

$$\langle n \rangle_R = \nu \frac{\Phi}{\Phi_0} , \tag{5.11}$$

where $\Phi$ is the magnetic flux enclosed inside the loop and $\Phi_0 = \frac{hc}{e}$ is the quantum flux unit. The validity of the relation (5.11) can be verified as

$$\Phi = BS \quad , \quad \Phi_0 = \frac{\hbar c 2\pi}{e} \quad , \quad \frac{\hbar c}{eB} = l_0^2 \quad , \tag{5.12}$$

and by using these relations we get

$$\nu \frac{\Phi}{\Phi_0} = \nu BS \frac{e}{2\pi\hbar c} = \nu S/(2\pi l_0^2) \quad . \tag{5.13}$$

Substituting in (5.13) the relation $\nu/(2\pi l_0^2) = n_0$ from (5.3) we get (5.11) where $\langle n \rangle_R = n_0 S$.

The result (5.10) for the geometric phase will be in agreement with a straight forward AB calculation only if we exchange the electron charge $e$ by an effective electron charge $e^*$ as

$$\gamma(C) = \frac{e^*}{\hbar c}\oint \vec{A}\cdot d\vec{l} = 2\pi \frac{e^*}{e}\frac{\Phi}{\Phi_0} \tag{5.14}$$

By comparing Eqs. (5.14) and (5.10) we find for the charge of the quasihole

$$e^* = -\nu e \tag{5.15}$$

This is only one example of obtaining *fractional charge* from geometric phase calculations [52,53]. Fractional charges have been obtained from the Hall conductance experiments [1,2] and have been related to quantization of the Hall effect. In any case the concept of Anyons in quantum Hall effects are related to fractional electron charges and topological geometric phases' calculations.

## 6. Summary

Various phenomena in Quantum Mechanics which are related to geometric phases have been reviewed and explained.. The present article is an extension of a previous Review in order to include more fields and some of the new developments in quantum mechanical topological effects.

The AB effect, $U(1)$ and $SU(2)$ phase rotations have been related to the basic topological concepts of 'parralelism', 'connections' and 'curvatures'. The theory of these effects have been presented in such way that the analogy with other topological quantum effects will be obvious. It has been shown that there are two alternative approaches to Schrodinger equation topological effects which usually give equivalent results. However the approach by which the wave function is single-valued can lead to interesting results.

Implementation of quantum gates by the use of Holonomic Quantum Computation (HQC) where information is encoded in degenerate states has been reviewed. The use of this method with the Wilczec-Zee connections has been demonstrated by applying 'dark states' for implementing the $CNOT$ gate.

The 'Anyons' concept has been related to the symmetries of the wave functions in a two-dimensional space. Anyons' effects including fractional charges and their relation to geometric phases has been demonstrated by analyzing in an example certain Quantum Hall effects.

**References**

[1] A. Shapere and F. Wilczec, *Geometric Phases in Physics*, (World Scientific, Singapore, 1989).

[2] A. Bohm, A. Mostafazadeh, H. Koizumi, Q. Niu and J. Zwanziger, *The geometric Phase in Quantum Systems* (Springer, Berlin, 2003).

[3] M. Nakahara, R. Rahimi, and A. SaiToh, eds. *Mathematical Aspects of Quantum Computing 2007* (World Scientific, Singapore, 2008).

[4] Y. Ben-Aryeh, J. Opt. B: Quantum Semiclass. Opt. **6** R1-R18 (2004).

[5] M.V. Berry, Proc. R. Soc. Lond. A **392** 45 (1984).

[6]   C.Nash and S.Sen, *Topology and Geometry for Physicists* (Academic Press, London, 1983).

[7]   Y.Aharonov and D.Bohm, Phys.Rev. **115** 485 (1959).

[8]   C.A. Mead, Rev.Mod. Phys. **64** 51 (1992).

[9]   I.J.R. Aitchison, Physica Scripta **T23** 12 (1988).

[10]  M.A. Nielsen and I.L. Chuang, *Quantum Computation and Quantum Information* (Cambridge Press, Cambridge, 2001).

[11]  G.Benenti, G.Casati, and G.Strini, *Principles of Quantum Computation and Information*, Volumes 1, 2 (World Scientific, Singapore, 2005).

[12]  V.Vedral, Int. J. Quantum Information **1** 1 (2003).

[13]  F. Wilczek and A. Zee, Phys. Rev. Lett. **52** 2111 (1984).

[14]  A. Zee, Phys.Rev.A **38** 1 (1988).

[15]  J.Pachos, P.Zanardi and M.Rasetti, Phys.Rev. A **61** 010305(R) (1999).

[16]  P.Zanardi and M.Rasetti, Phys. Lett.A **264** 94 (1999).

[17]  J.Pachos and S.Chountasis, Phys.RevA **62** 052318 (2000).

[18]  A.Recati, T.Calarco, P.Zanrdi, J.I. Cirac and P.Zoller, Phys.Rev.A **66** 032309 (2002).

[19]  A.O. Niskanen, M.Nakahara and M.M. Salommaa, Quantum information and Computation **2** 560 (2002).

[20]  S.Tanimura, "Holonomic Quantum Computing and its Optimization" in *Mathematical Aspects of Quantum Computing 2007*, M.Nakahara, R.Rahimi, and A.SaiToh, eds.(World Scientific, Singapore 2008).

[21]  S.Tanimura, D. Hayashi and M.Nakahara, Phys. Lett.A **325** 199 (2004).

[22]  I.Fuentes-Guridi, F. Girelli and E.Livine, Phys. Rev.Lett. **94** 020503 (2005).

[23]  L.-A Wu,  P. Zanardi and D.A. Lidar,  Phys.Rev.Lett. **95**  130501 (2005).

[24]  G. Florio, Open Sys.& Information Dyn.  **13**  263 (2006).

[25]  J.M. Leinaas and J. Myrheim,  Nuovo Cimento  **37**  1  (1977).

[26]  J.Myrheim, "Anyons" in *Topological Aspects of Low Dimensional Systems,* Nato Advanced   Study , Les  Houches,  A.Comtet,  T. Jolicoeur ,  S.Quavry and   F.David eds. (Springer , Berlin, 1998).

[27]  C.Nayak,  S.H.Simon,  A.Stern,  M.Freedman and  S.D.Sarma,  Rev.Mod. Physics  **80**  1083  (2008).

[28]  G.K.Brennen and   J.K. Pacho*s*,  Proc.Roy.Soc.A  **464**  1  (2008).

[29]   D.Yoshioka, *The quantum Hall Effect*  (Springer, Berlin, 1998).

[30]   T.Chackraborty and P. Pietilainen,  *The Quantum Hall effects*  (Springer, Berlin, 1995).

[31]  R.E.Prange  and   S.M. Girvin eds. *The Quantum Hall effect*, (Springer,  Singapore, 1987)

[32]  C. Hong-Mo and T.S. Tsun , *Some Elementary Gauge Theory Concepts* (World Scientific, Singapore , 1993).

[33]  Y.Ben-Aryeh and A.Mann , J.Opt.Soc.Am. B **17**  2052 (2000).

[34]  Y.Ahaonov and J.Anandan ,  Phys.Rev.Lett. **58** 1593  (1987).

[35]  S.Pancharatnam , The Proceeding of the Indian Academy of Sciences, Vol XLIV  247 (1956).

[36]  Y.Ben-Aryeh., J.Mod. Opt. **50**  2791 (2003).

[37]  J.Moody , A. Shapere and F.Wilczec , Phys.Rev.Lett..**56**  893 (1986).

[38]  Y.Ben-Aryeh ,  J.Mod. Opt. **49**  207   (2002).

[39]  Y.Ben-Aryeh, J.Mod. Opt. **54**  565   (2007).

[40]  Y.Ben-Aryeh, J.Opt.Soc.Am. B **25**  1965 (2008).

[41]  R.G. Unanyan,  B.W. Shore and K.Bergmann , Phys. Rev.A  **59**  2910 (1990).

[42]  B. Tregenna, A. Beige and P.L. Knight,  Phys. Rev.A  **65**  032305 (2002).

[43]  F.Wilczec,  Phys.Rev.Lett. **48**  1144 (1982).

[44]  B.Paredes, P.Fedichev, J.I.Cirac and P.Zoller , Phys. Rev.Lett.  **87**  010402 (2001).

[45]  R.M.Serra,  A.Carollo,  M.F.Santos and V.Vedral,  Phys.Rev.A  **70** 044102  (2004).

[46]  C.Weeks, G.Rosenberg, B.Seradjeh and M.Franz, Nature Physics **3**   796  (2007).

[47]  D.Arovas, J.R. Schrieffer and  F.Wilczec , Phys.Rev.Lett.**53**  722 (1984).

[48]  T. Ootsuka and  K. Sakuma, "Braid Group and Topological Quantum Computing"  in *Mathematical   Aspects  of  Quantum  Computing  2007* ,   M.Nakahara, R.Rahimi , and  A.SaiToh , eds. (World   Scientific, Singapore 2008).

[49]  R.B. Laughlin, Phys.Rev.Lett. **50**  1395 (1983).

[50]  R.B. Laughlin, Physic. Rev. B  **27**  3383 (1983).

[51]  R.B. Laughlin , "Elementay Theory:the Incompressible Quantum Fluid " in *The Quantum Hall effect*, eds R.E.Prange and  S.M. Girvin (Springer,  1987) p.233.

[52]  F.Wilczec , Phys.Rev.Lett. **49**  957 (1982).

[53]   D.P.Arovas, "Topics in Fractional Statistics" in  *Geometric  Phases  in  Physics*,  eds.A. Shapere and F. Wilczec (World  Scientific , Singapore, 1989 ) p. 284  .